\documentclass[aps,nofootinbib,preprintnumbers,amsmath]{revtex4-2}
\usepackage{graphicx, hyperref, xcolor}
\usepackage{subcaption}

\newcommand\pubdate{\today}

\def\Title#1{\begin{center} {\Large #1 } \end{center}}
\def\Author#1{\begin{center}{ \sc #1} \end{center}}
\def\Address#1{\begin{center}{ \it #1} \end{center}}

\newcommand\pubblock{\rightline{\begin{tabular}{l}  \\ 
         \pubdate  \end{tabular}}}
\newenvironment{Abstract}{\begin{quotation}  }{\end{quotation}}
\newenvironment{Presented}{\begin{quotation} \begin{center} 
             PRESENTED AT\end{center}\bigskip 
      \begin{center}\begin{large}}{\end{large}\end{center} \end{quotation}}

\def\eq#1{{Eq.~(\ref{#1})}}
\def\eqs#1{{Eqs.~(\ref{#1})}}

\newcommand{\ben}{\begin{eqnarray*}}
\newcommand{\een}{\end{eqnarray*}}
\newcommand{\un}[1]{\underline{#1}}

\newcommand{\tr}{\mbox{tr}}

\newcommand{\as}{\alpha_s}

\newcommand{\dhd}{{\textstyle d}
\lower.03ex\hbox{\kern-0.38em$^{\scriptstyle-}$}\kern-0.05em{}}
\newcommand{\dbar}{{\textstyle \delta}
\lower.03ex\hbox{\kern-0.38em$^{\scriptstyle-}$}\kern-0.05em{}}

\makeatletter
\DeclareRobustCommand{\cev}[1]{%
  {\mathpalette\do@cev{#1}}%
}
\newcommand{\do@cev}[2]{%
  \vbox{\offinterlineskip
    \sbox\z@{$\m@th#1 x$}%
    \ialign{##\cr
      \hidewidth\reflectbox{$\m@th#1\vec{}\mkern4mu$}\hidewidth\cr
      \noalign{\kern-\ht\z@}
      $\m@th#1#2$\cr
    }%
  }%
}
\makeatother

\newcommand{\iUV}{
    \int \displaylimits^z_{\frac{1}{s x_{10}^2}} \frac{dz^\prime }{z^\prime} \int \displaylimits^{x_{10}^2}_{\frac{1}{z^\prime s}} \frac{dx_{21}^2}{x_{21}^2}
}
\newcommand{\iIR}{
\int \displaylimits^{z}_{\frac{\Lambda^2}{s}} \frac{dz^\prime}{z^\prime} \int \displaylimits^{\mathrm{min}[\frac{z}{z^\prime} x_{10}^2, \frac{1}{\Lambda^2}]}_{\mathrm{max}[x_{10}^2, \frac{1}{z^{\prime}s}]} \frac{d x_{21}^2}{x_{21}^2}
}

\begin{document}
\pagenumbering{arabic}

\begin{titlepage}
 \pubblock
\vfill
\Title{Orbital Angular Momentum at Small $x$ Revisited}
\Author{Brandon Manley}
\Address{Department of Physics, The Ohio State University, Columbus, OH 43210, USA}
\begin{Abstract}
   We revisit the problem of the small Bjorken-$x$ asymptotics of the quark and gluon orbital angular momentum (OAM) distributions in the proton utilizing the revised formalism for small-$x$ helicity evolution derived recently in \cite{Cougoulic:2022gbk}. We relate the quark and gluon OAM distributions at small $x$ to the polarized dipole amplitudes and their (first) impact-parameter moments. To obtain the $x$-dependence of the OAM distributions, we derive novel small-$x$ evolution equations for the impact-parameter moments of the polarized dipole amplitudes in the double-logarithmic approximation (summing powers of $\as \ln^2(1/x)$ with $\as$ the strong coupling constant). We solve these evolution equations numerically and extract the large-$N_c$, small-$x$ asymptotics of the quark and gluon OAM distributions, which we determine to be
    \begin{align}
        \label{asym_res_abs}
        L_{q+\bar{q}}(x, Q^2) \sim L_{G}(x,Q^2) \sim \Delta \Sigma(x, Q^2) \sim \Delta G(x,Q^2) \sim \left(\frac{1}{x}\right)^{3.66 \, \sqrt{\frac{\as N_c}{2\pi}}},
    \end{align}
in agreement with \cite{Boussarie:2019icw} within the precision of our numerical evaluation (here $N_c$ is the number of quark colors). We also investigate the ratios of the quark and gluon OAM distributions to their helicity distribution counterparts in the small-$x$ region. 
\end{Abstract}
\begin{Presented}
DIS2023: XXX International Workshop on Deep-Inelastic Scattering and
Related Subjects, \\
Michigan State University, USA, 27-31 March 2023 \\
     \includegraphics[width=9cm]{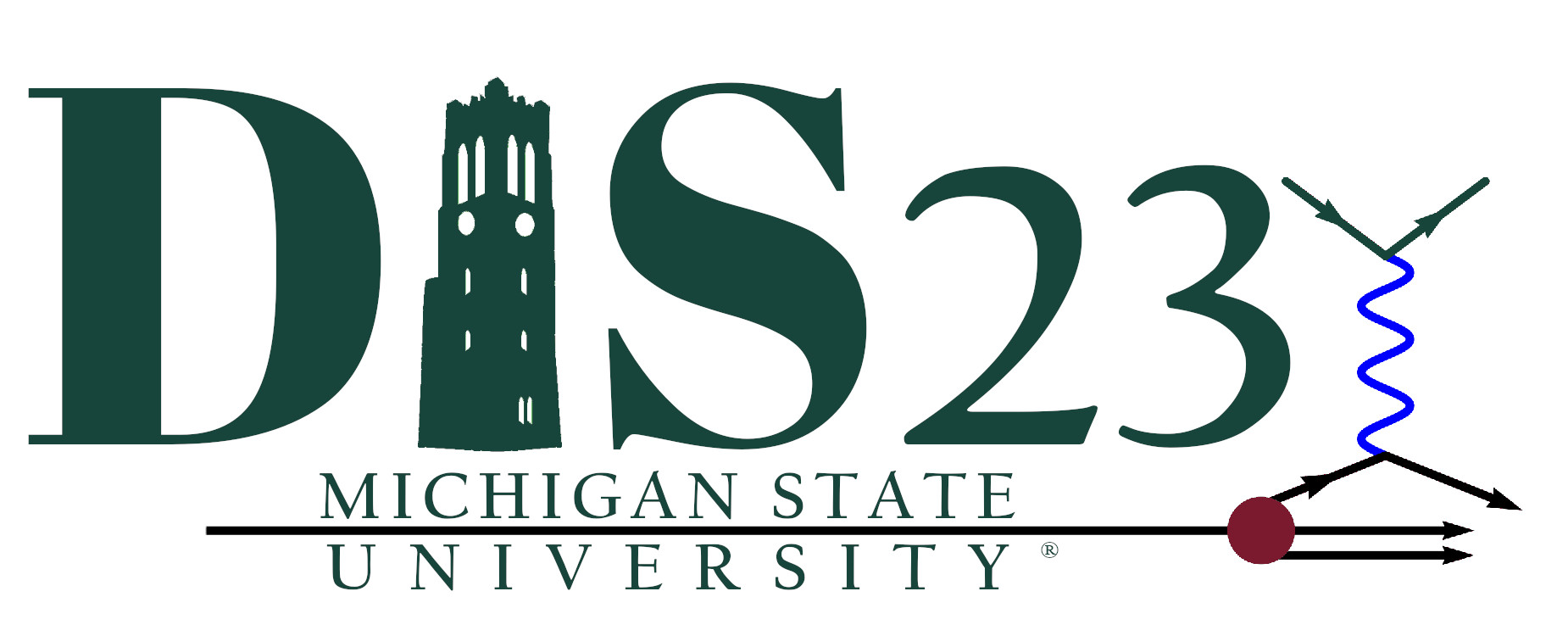}
\end{Presented}
\vfill
\end{titlepage}

\section{Introduction}
These proceedings are based on a forthcoming paper \cite{KovchegovManley}, where we present these findings in more detail. 
\newline
\indent The total spin of the proton comes from the sum of its constituents. The spin and orbital angular momenta of the constituents both contribute to the proton's spin. For example, here we consider the Jaffe-Manohar sum rule \cite{Jaffe:1989jz}, 
\begin{align}
    \label{spinsumrule}
    S_{q+\bar{q}}(Q^2) + L_{q+\bar{q}}(Q^2) + S_G(Q^2) + L_G(Q^2) = \frac{1}{2},
\end{align}
where $S_{q+\bar{q}}\, (S_G)$ is the quark (gluon) spin angular momentum and $L_{q+\bar{q}}\, (L_G)$ is the quark (gluon) orbital angular momentum, all of which depend on the renormalization scale $Q^2$. In this work, we mainly focus on the orbital angular momenta, which can be expressed as integrals over their distributions in the Bjorken-$x$ variable,
\begin{align}
    \label{oams}
    L_{q+\bar{q}}(Q^2) = \int \displaylimits^1_0 dx\, L_{q+\bar{q}}(x,Q^2),  \hspace{1cm}
    L_{G}(Q^2) = \int \displaylimits^1_0 dx\, L_{G}(x,Q^2). 
\end{align}
\indent Experimentally, it would require infinite energy to go to down to $x=0$ in the integrals of \eq{oams}. Therefore, we can only ever make measurements down to some $x_{\mathrm{min}} >0$, and theoretical input is needed to understand the region $x < x_{\mathrm{min}}$. This work intends to address this problem by determining the asymptotic behavior for $L_{q+\bar{q}}(x,Q^2)$ and $L_G(x,Q^2)$ as $x\to 0$ (but outside the saturation region). In the dipole formalism at small $x$, the asymptotic behavior of these distributions is related to the evolution of the polarized dipole amplitudes and their first impact-parameter moments. We introduce these polarized dipole amplitudes and their moments in section \ref{sec:oamdists}, discuss their evolution in section \ref{sec:evol}, and report some numerical results, including \eq{asym_res_abs}, obtained by solving these evolution equations in section \ref{sec:numres}. We conclude in section \ref{sec:con}.

\section{OAM Distributions at Small $x$}
\label{sec:oamdists}
The quark and gluon orbital angular momentum contributions to \eq{spinsumrule} can be expressed through the use of quantum Wigner functions $W_{q,G}$ \cite{Lorce:2011kd, Lorce:2011ni, Ji:2012ba, Kovchegov:2019rrz}, 
\begin{align}
    \label{oam_op}
    L_{q, G}(Q^2) = \int \frac{d^2 b_\perp db^- d^2 k_\perp dk^+}{(2\pi)^3} \left(\un b \times \un k\right) W_{q, G}(k,b),
\end{align}
where we use $\un x = (x^1, x^2)$ for the transverse variables. The orbital angular momentum distributions can then be obtained via \eq{oams}. The Wigner functions can be extracted from unpolarized quark and gluon transverse momentum dependent (TMD) distributions in a longitudinally polarized proton \cite{Mulders:1995dh, Kovchegov:2018znm, Kovchegov:2019rrz}. They are 
\begin{subequations}
    \label{Wigners}
\begin{align}
    W_{q+\bar{q}}^{\mathrm{SIDIS}}(k, b) &= 2 \sum_X \int dr^-\, d^2 r_\perp \, e^{i k\cdot r} \Bigg \langle \bar{\psi}\left( b - \frac{1}{2} r\right) V_{\un b - \frac{1}{2} \un r}\left[b^- - \frac{1}{2}r^-, \infty \right] |X\rangle 
    \\ \notag 
    & \hspace{4.5cm} \times 
    \left( \frac{\gamma^+}{2}\right) \langle X| V_{\un b + \frac{1}{2} \un r}\left[\infty, b^- + \frac{1}{2}r^-\right]
    \psi\left(b + \frac{1}{2}r \right)
    \Bigg \rangle, \\
    W_G^{\mathrm{dipole}}(k, b) &= \frac{4}{xP^+} \int dr^- \, d^2 r_\perp \, e^{i xP^+ r^- - i \un k \cdot \un r} 
    \\ \notag 
    & \hspace{0cm} \times \bigg\langle 
    \mathrm{tr}\left[
        F^{+i}\left(b- \frac{1}{2}r\right) \mathcal{U}^{[+]}\left[b - \frac{1}{2}r, b +\frac{1}{2}r \right] F^{+i}\left(b+ \frac{1}{2}r\right) \mathcal{U}^{[-]}\left[b + \frac{1}{2}r, b -\frac{1}{2}r \right]
    \right]
    \bigg \rangle,
\end{align}
\end{subequations}
where $k^+ = xP^+$, and we have averaged in the proton's wavefunction. The large brackets here indicate this averaging, as done in the Color Glass Condensate (CGC) formalism (see for example \cite{Iancu:2003xm}). Here, we use the following notation for the fundamental Wilson lines
\begin{align}
    V_{\un x}[b^-, a^-] = \mathcal{P} \exp \left[ ig  \int \displaylimits_{a^-}^{b^-} dx^- A^+(x^+=0, x^-, \un x)\right],
\end{align}
with $V{\un x} \equiv V_{\un x}[\infty, -\infty]$. We denote the future- and past- pointing Wilson line staples by $\mathcal{U}^{[+]}$ and $\mathcal{U}^{[-]}$ respectively. We have also chosen the gauge links to represent the semi-inclusive deep inelastic scattering (SIDIS) quark and dipole gluon distributions. 
\newline
\indent To describe the OAM distributions at small $x$, we need the polarized dipole amplitudes and their first impact-parameter moments. Consider a projectile-target system with a center of a mass energy $s$ with the target moving mostly along the plus light-cone direction and the projectile moving mostly along the minus light-cone direction. We can treat the projectile as a dipole with the transverse positions of the quark and antiquark denoted by either of $\un x_0, \un x_1$, with the softer parton having momentum fraction $z$. At the sub-eikonal level (one power of $x$ suppression), we can form two types of polarized dipole amplitudes \cite{Cougoulic:2022gbk},
\begin{subequations}
    \label{polamps}
\begin{align}
    \label{q10}
    Q_{10}(zs) &= \frac{zs}{2N_c} \mathrm{Re} \left\langle
    \mathrm{T\;\tr}\left[V_{\un x_0} \left(V_{\un x_1}^{\mathrm{pol}[1]} \right)^\dagger \right] + \mathrm{T\;\tr}\left[V_{\un x_1}^{\mathrm{pol}[1]} V_{\un x_0}^\dagger \right]
    \right\rangle,
    \\
    \label{g2}
    G^i_{10}(zs) &= \frac{zs}{2N_c} \mathrm{Re} \left\langle
    \mathrm{T\;\tr}\left[V_{\un x_0}^\dagger V_{\un x_1}^{i\,G[2]} \right] + \mathrm{T\;\tr}\left[\left(V_{\un x_1}^{i\,G[2]} \right)^\dagger V_{\un x_0} \right]
    \right\rangle.
\end{align}
\end{subequations}
$V_{\un x_1}^{\mathrm{pol}[1]}$ and $V_{\un x_1}^{i\,G[2]}$ are two different types of polarized propagators in the target shockwave \cite{Kovchegov:2015pbl}. Their operator definitions, along with the complete set of polarized propagators at sub-eikonal level, can be found in \cite{Cougoulic:2022gbk}. An earlier analysis \cite{Kovchegov:2019rrz} of the OAM distributions included only contributions from \eq{q10}. We aim here to correct that result by including contributions from \eq{g2} as well as \eq{q10}. 

Since \eqs{oam_op} involves integration over all impact-parameters, it is more convenient to work with the impact-parameter integrated quantities\footnote{There are other possible tensor structures on the right hand side of \eq{im-int-qs}, but they do not contribute to the OAM distributions, so we do not consider them here.},
\begin{subequations}
    \label{im-int-qs}
\begin{align}
    \int d^2 \left(\frac{\un x_0 + \un x_1}{2}\right) Q_{10} (zs) &= Q(x_{10}^2, zs), \\
    \int d^2 \left(\frac{\un x_0 + \un x_1}{2}\right) G^i_{10} (zs) &= \epsilon^{ij} x_{10}^j \, G_2(x_{10}^2,zs), \\
    \label{momI3}
    \int d^2 x_1 \, x_1^i Q_{10}(zs) &= x_{10}^i \, I_3(x_{10}^2,zs), \\
    \label{momI4}
    \int d^2 x_1 \, x_1^i G^j_{10}(zs) &= \epsilon^{ij} x_{10}^2 \, I_4(x_{10}^2,zs)  + \epsilon^{ik} x_{10}^k x_{10}^j \, I_5(x_{10}^2,zs) 
     + \epsilon^{jk} x_{10}^k x_{10}^i \, I_6(x_{10}^2,zs),
\end{align}
\end{subequations}
where $x_{10} \equiv |\un x_1 - \un x_0|$. We have also included the first $x_1$-moments in \eqs{momI3} and (\ref{momI4}), which also contribute to the OAM distributions. We parametrize these moments by the functions $I_3, I_4, I_5, I_6$, which we call the ``moment" amplitudes. 
\newline
\indent Using \eqs{Wigners} in \eq{oam_op}, expanding in $x$, and keeping only the leading terms, we arrive at the following small-$x$ expressions for the quark and gluon OAM distributions 
\begin{subequations}
    \label{smallxexp}
\begin{align}
    L_{q+\bar{q}}(x, Q^2) &= \frac{N_c N_f}{2\pi^3} \int \displaylimits^1_{\Lambda^2/s} \frac{dz}{z} \int \displaylimits^{\mathrm{min}\left[\frac{1}{zQ^2}, \frac{1}{\Lambda^2} \right]}_{\frac{1}{zs}} \frac{dx_{10}^2}{x_{10}^2} \left[Q - 3 \, G_2 - I_3 - 2\, I_4 + I_5 + 3\, I_6 \right](x_{10}^2, zs),
    \\
    L_G(x,Q^2) &= -\frac{2N_c}{\as \pi^2}\left[2\,I_4 + 3\, I_5 + I_6\right]\left(x_{10}^2 = \frac{1}{Q^2}, \, zs = \frac{Q^2}{x} \right),
\end{align}
\end{subequations}
where $\Lambda$ is an infrared (IR) cutoff. The expressions in \eqs{smallxexp} are valid in the double logarithmic approximation (DLA), summing powers of $\as \ln^2(1/x)$; we have discarded terms that are sub-leading in this approximation. Once we determine the asymptotic form of the dipole ($Q,G_2$) and moment ($I_3, I_4, I_5, I_6$) amplitudes, \eqs{smallxexp} will allow us to determine the asymptotic form of the OAM distributions.

\section{Moment and regular amplitude evolution equations in the large-$N_c$ limit}
\label{sec:evol}
The polarized dipole amplitudes in \eqs{polamps} obey integral evolution equations in the dipole size, $x_{10}^2$, and partonic center of mass energy $zs$ \cite{Cougoulic:2022gbk}. Although these equations do not close in general, they do close in the large-$N_c$ limit. In this case, the relevant evolution equations are given by Eqs. (118) for $G_{10}(zs)$\footnote{In the large-$N_c$ limit, we denote $Q_{10}(zs)$ by $G_{10}(zs)$.} and Eqs. (128) for $G^i_{10}(zs)$ of \cite{Cougoulic:2022gbk}.  
By integrating over impact parameters and keeping only terms that resum double logarithms of the form $\as \ln^2 \frac{1}{x}$, we get the DLA large-$N_c$ evolution equations for $G(x_{10}^2,zs)$ and $G_2(x_{10}^2, zs)$. These are derived in \cite{Cougoulic:2022gbk}, and are given by Eqs. (133) there. 
Similarly, we can start with Eqs.(118) and (128) of \cite{Cougoulic:2022gbk}, multiply both sides by $x_1^m$ and integrate both sides over $x_1$. By matching tensor structures and keeping only terms that resum double logarithms of energy, we arrive at the DLA large-$N_c$ evolution equations for $I_3, I_4, I_5$ and $I_6$,
    
\begin{align}
   \label{oamee}
    \begin{pmatrix}
    I_3 \\ 
    I_4 \\
    I_5 \\
    I_6 
    \end{pmatrix}(x_{10}^2, zs) &= \begin{pmatrix}
    I_3 \\ 
    I_4 \\
    I_5 \\
    I_6 
    \end{pmatrix}_0 (x_{10}^2, zs)  \\ \notag 
    & 
    + \frac{\as N_c}{4\pi} \iUV \,  \begin{pmatrix}
    2\,\Gamma_3 - 4\,\Gamma_4 + 2 \, \Gamma_5 + 6\, \Gamma_6 -2 \, \Gamma_2 \\ 
    0 \\
    0 \\
    0 
    \end{pmatrix}(x_{10}^2, x_{21}^2, z^\prime s) \\\notag &+ \frac{\as N_c}{4\pi} \iIR \, \begin{pmatrix}
        4 & -4 & 2 & 6 & -4 &  - 6 \\
        0 & 4 & 2 & -2 & 0 &  1 \\
        -2 & 2 & -1 & -3 & 2 &  3 \\
        0 & 0 & 0 & 0 & 2 &  4 \\
    \end{pmatrix} 
    \begin{pmatrix}
    I_3 \\ 
    I_4 \\
    I_5 \\
    I_6 \\ 
    G \\
    G_2
    \end{pmatrix} (x_{21}^2, z^\prime s).
\end{align}
The objects $\Gamma_3, \Gamma_4, \Gamma_5, \Gamma_6$ are the $x_1$-moments of the neighbor dipole amplitudes from helicity evolution, defined analogously to \eqs{im-int-qs}. 
The mixing between the regular and moment amplitudes in \eq{oamee} is reminiscent of the mixing between the helicity parton distribution functions (PDFs) and OAM distributions from polarized DGLAP evolution \cite{Hagler:1998kg, Hoodbhoy:1998}. 

\section{Numerical Results}
\label{sec:numres}
\begin{figure}[ht!]
     \centering
     \includegraphics[scale=0.6]{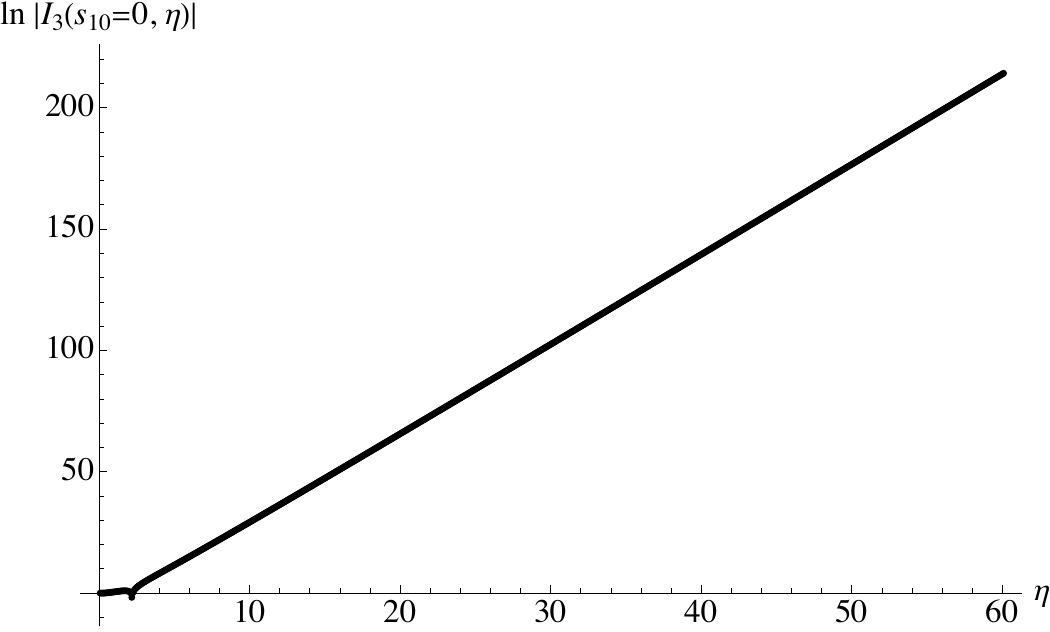}
     \caption{Plot of $\ln | I_3(s_{10}=0, \eta)|$ versus $\eta$ for $\Delta s_{10} = \Delta \eta \equiv \delta =0.075$. }
     \label{fig:I3}
 \end{figure}

Due to the complicated structure of \eq{oamee}, we solve it numerically here. We do this for a finite step size $\delta$ in the logarithmic variables $s_{10}= \sqrt{\frac{\as N_c}{2\pi}} \ln \frac{1}{x_{10}^2 \Lambda^2}, \, \eta = \sqrt{\frac{\as N_c}{2\pi}} \ln \frac{zs}{\Lambda^2}$. A logarithmic plot for one of the moment amplitudes\footnote{For the purposes of extracting the asymptotics, we only need the amplitudes at $s_{10}=0$.}, $I_3(s_{10}=0, \eta)$, is shown in Figure \ref{fig:I3}. 

 For large $\eta$, the amplitude seems to be exponential, prompting the following ansatz, 
\begin{align}
    \label{expansatz}
    I_3(s_{10}=0, \eta) \sim \exp\left[\alpha_{I_3} \eta \sqrt{\frac{2\pi}{\as N_c}} \right],
\end{align}
where $\alpha_{I_3}$ is the intercept. By regressing a linear model on $\ln|I_3(s_{10}=0, \eta)|$, we fit for the intercept $\alpha_{I_3}$. By repeating this procedure at different step sizes $\delta$ and maximum $\eta$ values, we can obtain a surface in the ($\delta, 1/\eta_{\mathrm{max}}$) space. We can then fit this surface to a polynomial model to extract the continuum limit ($\delta, 1/\eta_{\mathrm{max}}\to0$) intercept \cite{Kovchegov:2016weo, Cougoulic:2022gbk}.

\begin{figure}[ht!]
    \centering
    \begin{subfigure}[b]{0.48\textwidth}
         \centering
         \includegraphics[width=\textwidth]{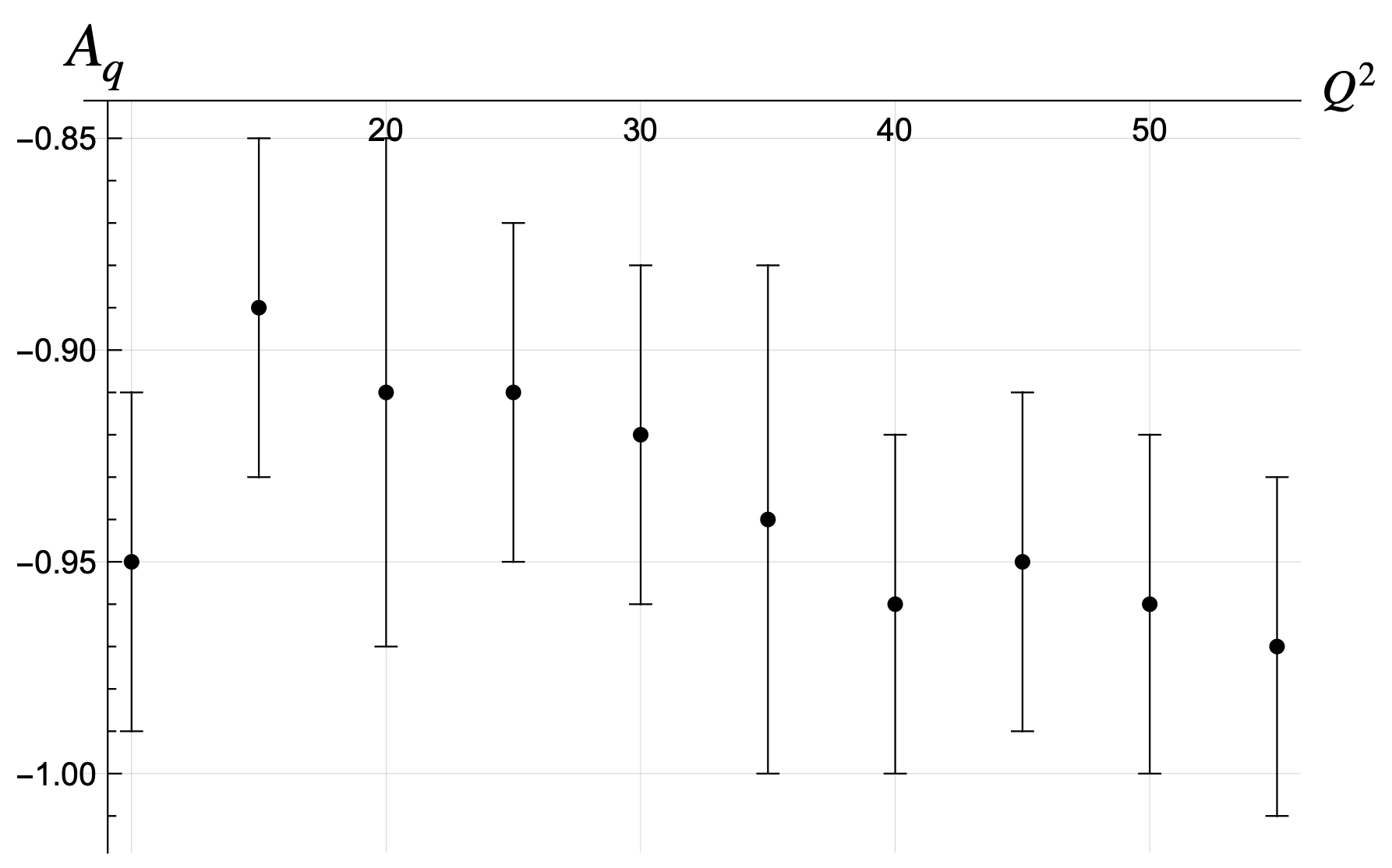}
         \caption{$A_q(Q^2)$}
     \end{subfigure}
     \begin{subfigure}[b]{0.48\textwidth}
         \centering
         \includegraphics[width=\textwidth]{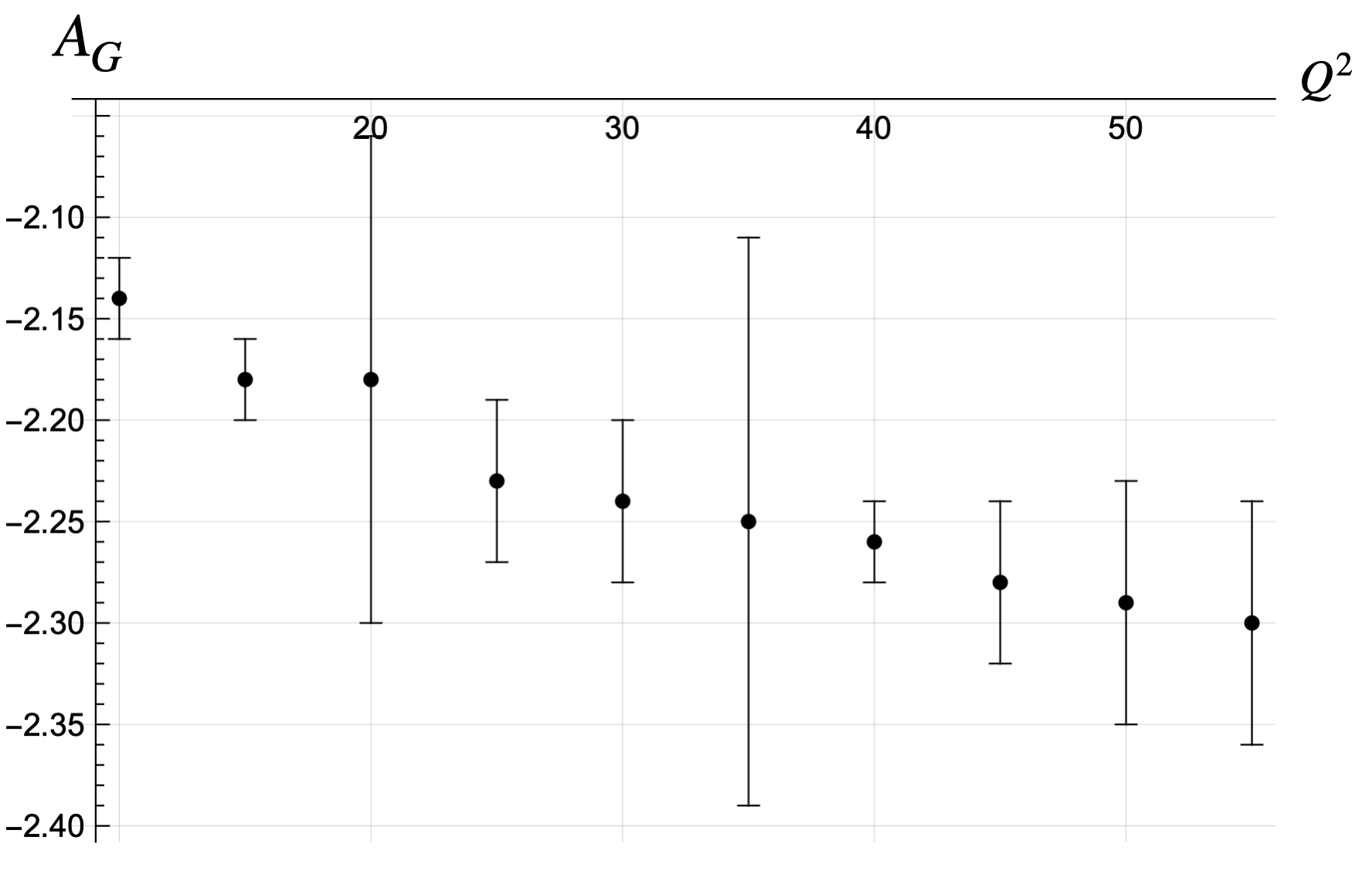}
         \caption{$A_G(Q^2)$}
     \end{subfigure}
    \caption{$A_q$ and $A_G$ as a function of $Q^2$ (given in values of $\mathrm{GeV}^2$).}
    \label{fig:ratios}
\end{figure}

\indent This procedure can be repeated for all of the amplitudes in \eqs{oamee}, as they all exhibit the exponential behavior of Figure \ref{fig:I3}. The resulting intercepts are all numerically consistent,
\begin{align}
    \label{intercepts}
    \alpha_{I_3} = \alpha_{I_4} = \alpha_{I_5} = \alpha_{I_6} = 3.66 \sqrt{\frac{\as N_c}{2\pi}}.
\end{align}
By use of \eq{expansatz} (and analogous ones for the rest of the amplitudes) and \eq{intercepts} in \eqs{smallxexp}, we arrive at the following small-$x$, large-$N_c$ asymptotics of the OAM distributions
\begin{align}
    \label{asym_res}
    L_{q+\bar{q}}(x, Q^2) \sim L_G(x,Q^2) \sim \left(\frac{1}{x} \right)^{3.66\sqrt{\frac{\as N_c}{2\pi}}}.
\end{align}
These are the same asymptotics as the corresponding helicity PDFs, $\Delta \Sigma(x,Q^2)$, $\Delta G(x,Q^2)$ and the $g_1$ structure function \cite{Cougoulic:2022gbk}. Therefore, \eq{asym_res} is an important indication that the OAM distributions are not suppressed relative to the helicity distributions at small $x$. 
\newline \indent Beyond the small-$x$ asymptotics, using \eqs{smallxexp}, we can also determine the ratio of the OAM distributions to the helicity PDFs. By plugging the numerical solution of \eqs{oamee} into \eqs{smallxexp}, we can calculate the OAM distributions numerically. Relations similar to \eqs{smallxexp} hold for the helicity PDFs as well. Through an analysis of the logarithmic plots of the OAM and helicity distributions as well as their derivatives, one can motivate the following ansätze for the distributions at small $x$,
\begin{align}
    \label{dist_ans}
    \ln |f(x,Q^2)| \approx \alpha_f \ln \frac{1}{x} + \beta_f + \frac{\delta_f}{\ln \frac{1}{x}} + \gamma_f \ln\left(\ln \frac{1}{x}\right),
\end{align}
where $f \in \{L_{q+\bar{q}}, L_G, \Delta \Sigma, \Delta G\}$. We can then approximate the ratios of the OAM to helicity distributions through the parameters $\alpha, \beta, \delta, \gamma$ for each $f$. Above, we found $\alpha_{L_{q+\bar{q}}} = \alpha_{\Delta \Sigma} = \alpha_{L_G} = \alpha_{\Delta G}$ and  numerically, one can show that $\gamma_{L_{q+\bar{q}}} = \gamma_{\Delta \Sigma} = \gamma_{L_G} = \gamma_{\Delta G}$. This means that the OAM to helicity PDF ratios are completely determined by the $\beta$ and $\delta$ parameters 
\begin{subequations}
    \label{ratio_ans}
    \begin{align}
            \frac{L_{q+\bar{q}}(x,Q^2)}{\Delta \Sigma(x,Q^2)} \approx - \exp\left(\beta_{L_{q+\bar{q}}} - \beta_{\Delta \Sigma}  \right)\left(1+ \frac{\delta_{L_{q+\bar{q}}} - \delta_{\Delta \Sigma}}{\ln \frac{1}{x}} \right) \equiv A_q + \frac{B_q}{\ln \frac{1}{x}}, 
            \\
            \frac{L_{G}(x,Q^2)}{\Delta G(x,Q^2)} \approx - \exp\left(\beta_{L_G} - \beta_{\Delta G}  \right)\left(1+ \frac{\delta_{L_G} - \delta_{\Delta G}}{\ln \frac{1}{x}} \right) \equiv A_G + \frac{B_G}{\ln \frac{1}{x}}.
    \end{align}
\end{subequations}
\eq{dist_ans} and  \eqs{ratio_ans} are valid up to $\mathcal{O}(\ln^{-2}(1/x))$ corrections. In general, $A_q, B_q, A_G, B_G$ depend on the renormalization scale $Q^2$. By performing a similar numerical procedure for the continuum limit outlined above for the intercepts, we can extract these parameters for a given $Q^2$. The constant term values, $A_q$ and $A_G$, are given for a range of $Q^2$ in Figure \ref{fig:ratios}.

\indent We can compare these to the results obtained in \cite{Boussarie:2019icw}. For helicity PDFs obeying the asymptotics $\Delta \Sigma (x,Q^2), \Delta G(x,Q^2) \sim x^{-\alpha}$, they predict in Eq. (7) of \cite{Boussarie:2019icw},
\begin{subequations}
    \label{hattaPred}
\begin{align}
    L_{q+\bar{q}}(x,Q^2) &= -\frac{1}{1+\alpha} \Delta \Sigma(x,Q^2) \approx - \Delta \Sigma(x,Q^2), \\
    L_G(x,Q^2) &= -\frac{2}{1+ \alpha}\Delta G(x,Q^2) \approx -2 \, \Delta G(x,Q^2),
\end{align}
\end{subequations}
 where, in the last equality, we have used $\alpha \sim \sqrt{\as}$ and expanded in $\sqrt{\as}$. Comparing to Fig. \ref{fig:ratios}, we observe, although the values are close, they are not entirely consistent. More work is needed to understand the difference between the two results and will likely require an analytic solution of \eqs{oamee}.

\section{Conclusions and Outlook}
\label{sec:con}
In this work, we have studied the OAM distributions at small $x$. We have revised and updated the results from \cite{Kovchegov:2019rrz} by including contributions from \eq{g2} in the calculation. We related the OAM distributions to the polarized dipole amplitudes and their first impact-parameter moments. We derived novel large-$N_c$ evolution equations for the latter, and solved them numerically to obtain the small-$x$, large-$N_c$ asymptotics of the OAM distributions. Importantly, within the precision of our numerics, we found the OAM distributions to have the same asymptotics as the quark and gluon helicity PDFs, as well as the $g_1$ structure function, in agreement with \cite{Boussarie:2019icw}. 
\newline 
\indent 
Furthermore, we studied the ratio of the OAM distributions to the helicity PDFs. This ratio in general depends on $x$ and renormalization scale. The result is close to the values predicted in \cite{Boussarie:2019icw} with some differences outlined above. 

\section*{Acknowledgments}
The author would like to thank Yuri Kovchegov for collaborating on this project, and Josh Tawabutr for helpful discussions and generously making his code for the helicity evolution available to him. 
This material is based upon work supported by
the U.S. Department of Energy, Office of Science, Office of Nuclear
Physics under Award Number DE-SC0004286. \\

\typeout{}
\bibliography{referencesn}

\providecommand{\href}[2]{#2}\begingroup\raggedright\begin{thebibliography}{10}

\bibitem{Cougoulic:2022gbk}
F.~Cougoulic, Y.V.~Kovchegov, A.~Tarasov and Y.~Tawabutr, \emph{Quark and gluon
  helicity evolution at small x: revised and updated},
  \href{https://doi.org/10.1007/jhep07(2022)095}{\emph{Journal of High Energy
  Physics} {\bfseries 2022} (2022) }.

\bibitem{Boussarie:2019icw}
R.~Boussarie, Y.~Hatta and F.~Yuan, \emph{{Proton Spin Structure at
  Small-$x$}},
  \href{https://doi.org/10.1016/j.physletb.2019.134817}{\emph{Phys. Lett.}
  {\bfseries B797} (2019) 134817}
  [\href{https://arxiv.org/abs/1904.02693}{{\ttfamily 1904.02693}}].

\bibitem{KovchegovManley}
Y.V.~Kovchegov and B.~Manley, \emph{Orbital angular momentum at small x
  revisited},  in preparation, 2023.

\bibitem{Jaffe:1989jz}
R.L.~Jaffe and A.~Manohar, \emph{{The G(1) Problem: Fact and Fantasy on the
  Spin of the Proton}},
  \href{https://doi.org/10.1016/0550-3213(90)90506-9}{\emph{Nucl. Phys.}
  {\bfseries B337} (1990) 509}.

\bibitem{Lorce:2011kd}
C.~Lorce and B.~Pasquini, \emph{{Quark Wigner Distributions and Orbital Angular
  Momentum}}, \href{https://doi.org/10.1103/PhysRevD.84.014015}{\emph{Phys.
  Rev.} {\bfseries D84} (2011) 014015}
  [\href{https://arxiv.org/abs/1106.0139}{{\ttfamily 1106.0139}}].

\bibitem{Lorce:2011ni}
C.~Lorce, B.~Pasquini, X.~Xiong and F.~Yuan, \emph{{The quark orbital angular
  momentum from Wigner distributions and light-cone wave functions}},
  \href{https://doi.org/10.1103/PhysRevD.85.114006}{\emph{Phys. Rev.}
  {\bfseries D85} (2012) 114006}
  [\href{https://arxiv.org/abs/1111.4827}{{\ttfamily 1111.4827}}].

\bibitem{Ji:2012ba}
X.~Ji, X.~Xiong and F.~Yuan, \emph{{Probing Parton Orbital Angular Momentum in
  Longitudinally Polarized Nucleon}},
  \href{https://doi.org/10.1103/PhysRevD.88.014041}{\emph{Phys. Rev.}
  {\bfseries D88} (2013) 014041}
  [\href{https://arxiv.org/abs/1207.5221}{{\ttfamily 1207.5221}}].

\bibitem{Kovchegov:2019rrz}
Y.V.~Kovchegov, \emph{{Orbital Angular Momentum at Small $x$}},
  \href{https://doi.org/10.1007/JHEP03(2019)174}{\emph{JHEP} {\bfseries 03}
  (2019) 174} [\href{https://arxiv.org/abs/1901.07453}{{\ttfamily
  1901.07453}}].

\bibitem{Mulders:1995dh}
P.J.~Mulders and R.D.~Tangerman, \emph{{The Complete tree level result up to
  order 1/Q for polarized deep inelastic leptoproduction}},
  \href{https://doi.org/10.1016/0550-3213(95)00632-X}{\emph{Nucl. Phys.}
  {\bfseries B461} (1996) 197}
  [\href{https://arxiv.org/abs/hep-ph/9510301}{{\ttfamily hep-ph/9510301}}].

\bibitem{Kovchegov:2018znm}
Y.V.~Kovchegov and M.D.~Sievert, \emph{{Small-$x$ Helicity Evolution: an
  Operator Treatment}},
  \href{https://doi.org/10.1103/PhysRevD.99.054032}{\emph{Phys. Rev.}
  {\bfseries D99} (2019) 054032}
  [\href{https://arxiv.org/abs/1808.09010}{{\ttfamily 1808.09010}}].

\bibitem{Iancu:2003xm}
E.~Iancu and R.~Venugopalan, \emph{{The Color glass condensate and high-energy
  scattering in QCD}},  in \emph{{Quark-gluon plasma 4}}, R.C.~Hwa and
  X.-N.~Wang, eds. (2003)
  [\href{https://arxiv.org/abs/hep-ph/0303204}{{\ttfamily hep-ph/0303204}}].

\bibitem{Kovchegov:2015pbl}
Y.V.~Kovchegov, D.~Pitonyak and M.D.~Sievert, \emph{{Helicity Evolution at
  Small-x}}, \href{https://doi.org/10.1007/JHEP01(2016)072}{\emph{JHEP}
  {\bfseries 01} (2016) 072}
  [\href{https://arxiv.org/abs/1511.06737}{{\ttfamily 1511.06737}}].

\bibitem{Hagler:1998kg}
P.~Hagler and A.~Schafer, \emph{{Evolution equations for higher moments of
  angular momentum distributions}},
  \href{https://doi.org/10.1016/S0370-2693(98)00414-6}{\emph{Phys. Lett.}
  {\bfseries B430} (1998) 179}
  [\href{https://arxiv.org/abs/hep-ph/9802362}{{\ttfamily hep-ph/9802362}}].

\bibitem{Hoodbhoy:1998}
P.~Hoodbhoy, X.~Ji and W.~Lu, \emph{Quark orbital-angular-momentum distribution
  in the nucleon},
  \href{https://doi.org/10.1103/physrevd.59.014013}{\emph{Physical Review D}
  {\bfseries 59} (1998) }.

\bibitem{Kovchegov:2016weo}
Y.V.~Kovchegov, D.~Pitonyak and M.D.~Sievert, \emph{{Small-$x$ asymptotics of
  the quark helicity distribution}},
  \href{https://doi.org/10.1103/PhysRevLett.118.052001}{\emph{Phys. Rev. Lett.}
  {\bfseries 118} (2017) 052001}
  [\href{https://arxiv.org/abs/1610.06188}{{\ttfamily 1610.06188}}].

\end{thebibliography}\endgroup
\bibliographystyle{JHEP}

\end{document}